\documentclass[twocolumn,preprintnumbers,amsmath,amssymb,prl,showpacs]{revtex4}

\usepackage{graphicx}
\usepackage{dcolumn}
\usepackage{bm}
\usepackage{rotating}

\begin{document}

\title{Quantum eigenstates and (pseudo-)classical accelerator modes \protect\\of the Delta-Kicked Accelerator}

\author{Gabriel Lemari\'e}
\affiliation{D\'epartement de Physique, E.N.S. de Cachan, 61 av du Pr\'esident Wilson, 94230 Cachan, France}
\author{Keith Burnett}
 \affiliation{Clarendon Laboratory, Department of Physics, University of Oxford, Oxford OX1 3PU, United Kingdom}
\date{\today}

\begin{abstract}
The quantum dynamics of a periodically driven system, the delta-kicked accelerator, is investigated in the semiclassical and pseudo-classical regimes~\cite{FGR}, where quantum accelerator modes are observed. We construct the evolution operator of this classically chaotic system explicitly. If certain quantum resonance conditions are fullfilled, we show one can reduce the evolution operator to a finite matrix, whose eigenvectors are the quasi-eigenstates. These are represented by their Husimi functions. In so doing, we are able to directly compare the pure quantum states with the classical states. In the semiclassical regime, the quantum states are found to be related to the classical KAM tori and the classical accelerator modes. In contrast, the quasi-eigenstates do not lie on the $\epsilon$-classical trajectories in the pseudo-classical regime. This shows a clear and important distinction between semiclassicality and the new type of pseudo-classicality found by Fishman, Guarneri and Rebuzzini~\cite{FGR}.       
\end{abstract}
\pacs{03.65.Ge, 32.80.Lg, 05.45.Mt }

\maketitle

In the present letter, we study the quantum eigenstates of a periodically driven system, the delta-kicked accelerator, which is a variant of the extensively studied kicked rotor (see \cite{Casati}). With the kicked rotor, particles are kicked by a sinusoidal potential whereas in the case of the delta-kicked accelerator, the kicked particles also experience the effect of gravity. Because of the explicit presence of gravity, the problem of finding the quantum states is closely related to that for the electrons in a lattice, in the presence of a homogeneous electric field (Wanier-Stark states \cite{Korsch}). The case of the kicked rotor is rather well understood: in particular we know that in the semiclassical regime, the eigenstates lie on classical regular trajectories (KAM tori). To date there has yet not been a corresponding investigation of the quantum states of the delta-kicked accelerator. In this paper, we will present evidence that the eigenstates of the delta-kicked accelerator lie on classical KAM tori in the semiclassical regime.\\
\indent Determining the quantum states of the delta-kicked accelerator is of particular interest as it enables us to examine the nature of new pseudo-classical regimes revealed by the appearance of quantum accelerator modes. The experimental realisation of the system, using cold atoms (see \cite{oxford}) lead to the observation of quantum accelerator mode: the momentum (measured in the falling frame) of a fraction of the falling atoms was found to increase (or decrease) linearly with the number of laser kicks. Accelerator modes appear in regimes which can be far from the semiclassical limit, even though they have similar features to those of the classical accelerator modes~\cite{FGR}. In fact Fishman and al. have shown that when the kicking periodicity is close to particular resonant times, the quantum dynamics is well modeled by a pseudo classical map and the behaviour of the quantum system is said to be ``$\epsilon$-classical''.\\
\indent The question then arises as to whether the eigenstates of the system are related to any $\epsilon$-classical trajectories. In other words, are the pseudo-classical and the semiclassical characteristics the same and what is the nature of these new pseudo-classical regimes? In order to answer these fundamental questions, we have to determine the full quantum states of the system. In the following, we will show how to obtain these states without approximation.\\
\indent The dynamics of the delta-kicked accelerator is defined by the following Hamiltonian:
\begin{eqnarray}
\hat{H}&=& \frac{\hat{p}^2}{2 m } + m g \hat{z}- \phi _d \; \cos(G \hat{z})\;\sum_l \delta (t- l T) \;, \label{Hamiltonian}
\end{eqnarray}
where $\phi_d$ is the strength of the kicking potential and $2 \pi/G$ its period. The positive direction of $z$ is taken to be the opposite to that of the acceleration due to gravity. 
\\ \indent As the above Hamiltonian is periodic in time, Floquet's theorem tells us that the solutions of the associated Schr\"odinger equation have the following form: $\vert \Psi (t)\rangle = e^{- i \omega t} \; \vert \phi _{\omega} (t)\rangle$, where $\omega$ is termed the quasi-energy and $\vert \phi _{\omega} (t)\rangle$ the quasi-eigenstate which is $T$-periodic in time. These are obtained by calculating the eigenvalues and eigenvectors of the time evolution operator. We can obtain this operator by standard methods. An elementary calculation gives its momentum representation in the form:
\begin{eqnarray}
\mathcal G(p_b, p_a, T) = \frac{1}{\hbar G} \; e^{  -\frac{i}{\hbar} \left(\frac{m g^2 T^3}{6}  + \frac{p_a^2 T}{2 m} - \frac{p_a g T^2}{2}\right) } \times\qquad \quad  \nonumber\\
 \times \frac{1}{2 \pi}  \int d \theta \; e^{ -i \theta \frac{mgT - p_a + p_b}{\hbar G} } \; e^{i k \cos(\theta)}  \; ,\;
\end{eqnarray}
where $k=\phi_d/\hbar$ and $\theta = G z \; [2\pi]$. Because $ e^{i k \cos(\theta)}$ is $2 \pi $ periodic, $(mgT - p_a + p_b)/{\hbar G}  $ must be an integer $q$, i.e.: $p_b = p_a -m g T + q \hbar G$. Then a simple recurrence argument shows that any momentum component present can be expressed in the following form: $p= q \hbar G - n_p \; m g T + \beta $, where $\beta$ is the so-called quasi-momentum. \\
\indent In the following, we will investigate the dynamics of the delta-kicked accelerator given that the parameters $T$, $\Omega = g G T^2/2 \pi$ and $\beta$ satisfy the conditions:
\begin{eqnarray}
T&=& \frac{M}{N} \; T_{1/2} = \frac{M}{N} \; \frac{2 \pi m}{\hbar G^2} \; , \\
\Omega &=& \frac{R}{S} \; ,\; \\
\beta &=& \hbar G  \left( \dfrac{l}{ M/N}+1 \right)\; , 
\end{eqnarray}
with $N$ taken as an even integer. This choice is crucial as it gives rise to quantum resonances (see \cite{articlelong}). It results in the set of momenta $q \hbar G - n_p \;  m g T + \beta$ to be discrete since:
\begin{eqnarray}
mgT =  \hbar G \frac{R}{S}\frac{N}{M} \; .
\end{eqnarray}
In turn, this makes it possible to express the evolution operator in terms of a finite matrix, using the discrete-ladder of momenta $\vert p_{q} = q \frac{\hbar G}{M S} + \beta \rangle$. The wave function is scaled on this discrete set of momenta, resulting in a vector whose components are: $\psi(q')= \langle p_{q'} \vert \psi   \rangle$. We make the following definition: $ \mathcal J_{q'-q}(k)= \frac{1}{2 \pi} \;\int d\theta \; e^{-i \theta \frac{R N + q - q'}{S M}} \; e^{i k \cos(\theta)}$, and hence the following matrix relation: 
$\psi(q,T)= \sum_{q'}  \mathcal U   _{q q' }  \psi(q',0) $, where: 
\begin{eqnarray}
\mathcal U _{q q'}&=& \hbar G \;  \mathcal G (p_{q}, p_{q'}, T)  \nonumber\\
&=&  e^{ -i \pi \left( q'^2 \frac{1}{{S}^2 M N} +2 q' \left( \frac{l}{S M} + \frac{1}{S N}\right) - q' \frac{R}{{S}^2 M}\right) } \; \mathcal J_{q'-q}(k)\; . \quad
\end{eqnarray}
It is straightforward to see that the matrix $\mathcal U$ obeys the following symmetry (recall that $N$ is taken as even): 
\begin{eqnarray}
\mathcal U _{q+ (N {S}^2 M), q'+ (N {S}^2 M)}= \mathcal U _{q q'} \; .
\end{eqnarray}
Hence, our ``stroboscopic'' representation of the dynamics is invariant under the momentum-eigenvalue translation:
$ q \hookrightarrow q + N S^2 M$. Floquet's theorem then tells us that the quasi-eigenstates of our system have the general form:
\begin{eqnarray}
\Phi_{s+ (N {S}^2 M) \nu }= e^{- i \theta_0 \nu} \; [\Phi(\theta_0)]_s \; , \quad 0\leq\theta_0\leq2 \pi \; ,
\end{eqnarray}
where now $ [\Phi(\theta_0)]_s$ is the eigenvector of the finite $N {S}^2 M \times N {S}^2 M$ matrix $[\mathcal U (\theta_0) ] $ defined by:
\begin{eqnarray}
[\mathcal U (\theta_0) ]_{s,s'} = \sum_{\nu'} \mathcal U_{s, s' + (N {S}^2 M) \nu'} e^{- i \theta_0 \nu'} \; .
\end{eqnarray}
Elementary calculations, and the use of the Poisson summation formula $\sum_{\nu} e^{i 2 \pi \nu \phi} = \sum_{\mu} \delta(\phi - \mu)$, make it possible to express the elements of the matrix $[\mathcal U (\theta_0) ]$. Thus:
\begin{eqnarray}
[\mathcal U (\theta_0) ]_{s,s'} = \qquad\qquad\qquad\qquad\qquad\qquad\qquad\qquad \; \;\nonumber\\
= e^{ -i \pi \left( s'^2 \frac{1}{{S}^2 M N} +2 s' \left( \frac{l}{{S} M} + \frac{1}{{S} N}\right) - s' \frac{{R}}{{S}^2 M}\right) } \; e^ {- i \theta_0 \frac{{R} N + s - s'}{{S}^2 M N} } \times \nonumber\\ 
\times  \frac{1}{{S}^2 N M}\; 
\sum_{\mu=1}^{{S}^2 N M} e^{ ik \cos\left(\frac{\theta_0}{{S} N}+ \frac{2 \pi \mu}{{S} N}\right) } \; 
e^{ -i 2 \pi\mu \frac{{R} N + s - s'}{{S}^2 M N}}  \; , \quad
\end{eqnarray}
The eigenvectors of $[\mathcal U (\theta_0) ] $ are found numerically, and used to determine the quasi-eigenstates in momentum representation $\left(\Phi_q \right)_q$.
\begin{figure}
\begin{center}
\includegraphics[width=3.7cm,height=3.7cm,angle=0]{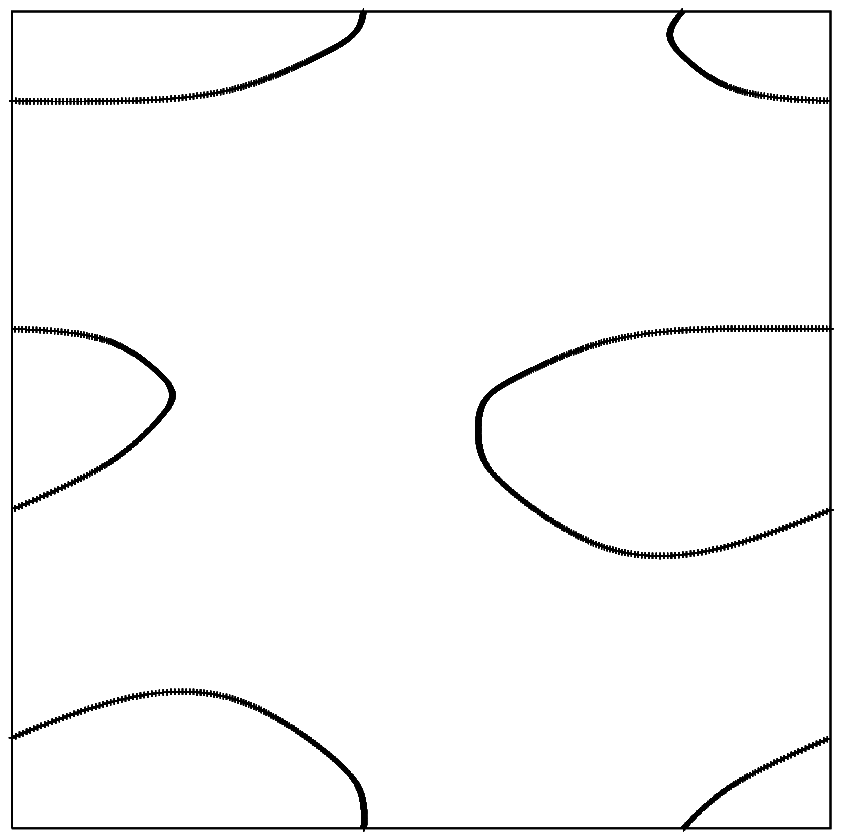}
\put(-58,108){$(a)$}
\put(-23,47){$B$}
\put(-88,6){$A'$}
\put(-88,100){$A$}
\includegraphics[width=3.7cm,height=3.7cm,angle=0]{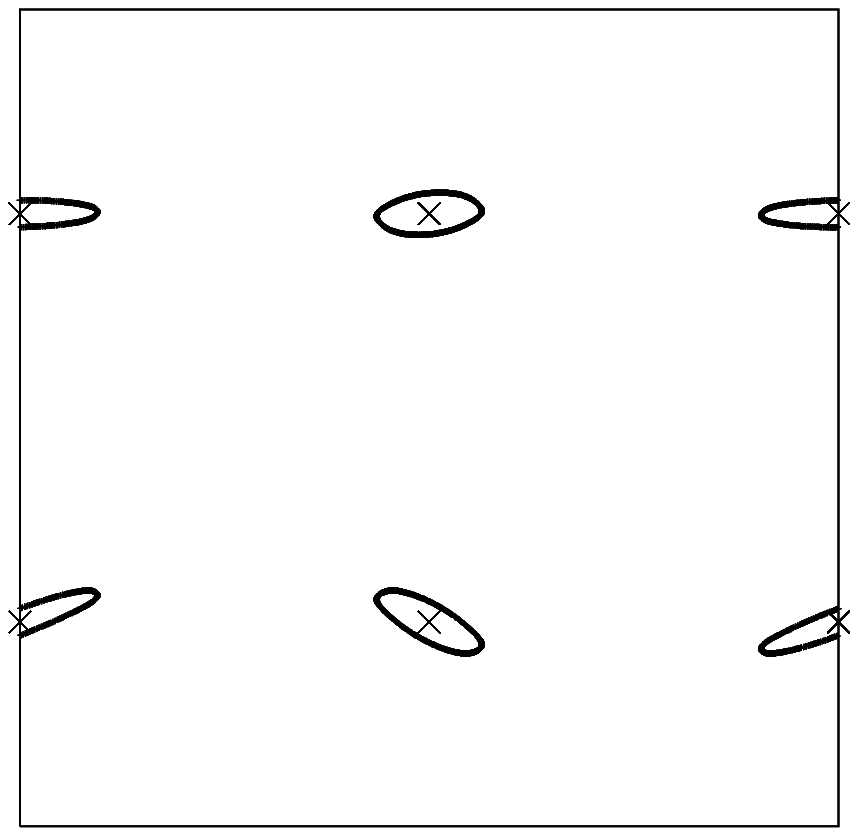}
\put(-58,108){$(b)$}
\end{center}
\begin{center}
\includegraphics[width=4cm,height=4cm,angle=0]{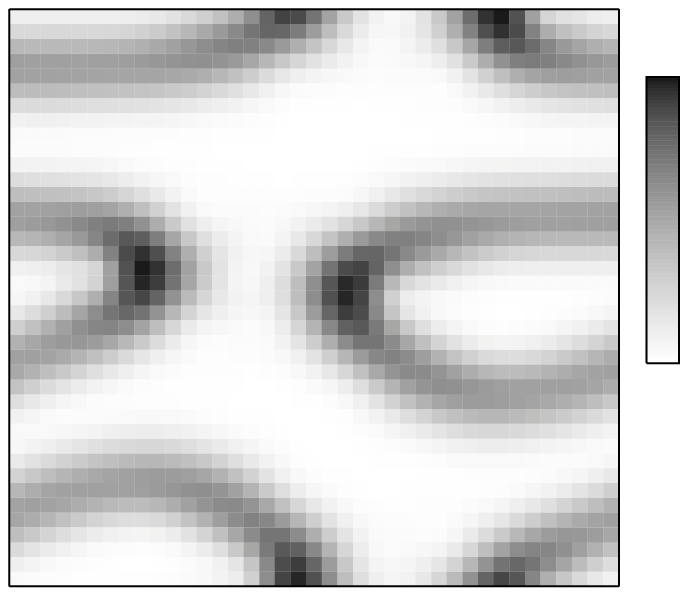}
\put(-105,55){$ \mathcal J$}
\put(-103,18){$0$}
\put(-107,92){$2 \pi$}
\put(-60,11){$\theta$}
\put(-97,11){$0$}
\put(-22,11){$2 \pi$}
\put(-60,102){$(c)$}
\put(-8,47){$0$}
\put(-8,83){$0.6$}
\put(-26,100){$|\langle \Phi \vert \theta , \mathcal J \rangle|^2 $}
\includegraphics[width=4cm,height=4cm,angle=0]{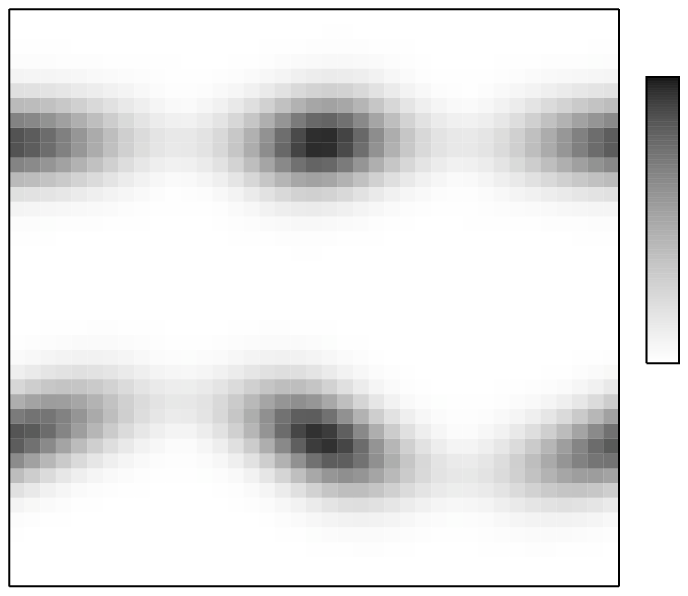}
\put(-60,102){$(d)$}
\put(-8,47){$0$}
\put(-8,83){$0.9$}
\caption{\small{Some classical and quantum states (in the semi-classical regime) of the delta-kicked accelerator. The parameters used are $\Omega=\frac{1}{2}$, $k=5$ and $T=\frac{1}{80}T_{1/2}$. $(a)$ and $(b)$: Poincar\'e sections (i.e. stroboscopic representations) of classical trajectories:  $(a)$ the KAM tori are generated by the initial condition $(\theta, \mathcal J) = (2.7, 0)$, $(b)$ the period-4 orbit is generated by the initial point $(\pi, \frac{\pi}{2})$. $(c)$ and $(d)$: The Husimi functions of two quantum quasi-eigenstates associated with different quasi-energies and for $\theta_0=0$. The quantum state in $(c)$ is closely related to the KAM tori in $(a)$. The state represented in $(d)$ lie on the period-4 orbit in $(b)$.}}
\label{q-cla}
\end{center}
\end{figure} 
\\ \indent As part of our investigation of the quantum-classical correspondence, we use a phase-space representation of the quantum dynamics. This representation makes it easier to relate the pure quantum states to the relevant classical ones. In order to have a phase-space representation of quantum dynamics, Gaussian smoothed Wigner functions (i.e. Husimi functions) are often used. In like manner we shall use coherent states to compute the Husimi functions of the quasi-eigenstates of the delta-kicked accelerator \cite{Chang}. The coherent states are minimum-uncertainty wave packets centred around their expectation values $(z',p')$ and they form a natural basis. The wave function of a quasi-eigenstate $\vert \Phi \rangle$ in such a representation is given by: $\langle \Phi \vert z', p' \rangle = \sum_q \langle \Phi \vert p_q \rangle \; \langle p_q \vert z', p' \rangle $, where:
\begin{eqnarray}
\langle p_q \vert z', p' \rangle = \left( \frac{1}{\pi \hbar \lambda}\right) ^{1/4} e^{ i \frac{(p' - 2 p_q) z'}{2 \hbar}} e^{-\left( \frac{p_q-p'}{\sqrt{2 \hbar\lambda}} \right) ^2 } \; ,
\end{eqnarray}
with $\lambda$ being a conveniently chosen parameter taken as $\lambda=m N/T_{1/2}$. One can show (see \cite{articlelong}) that the Husimi function of a quasi-eigenstate has the periodicity $z'\hookrightarrow z'+ M S \frac{2 \pi}{G}$ and $ p' \hookrightarrow p' + N S \hbar G$. This allows us to represent this function on a quantum phase-space cell  $\left[0, M S \frac{2 \pi}{G}\right] \times \left[0,\hbar G S N\right] $.
\\ \indent We shall now use the phase-space representation to investigate the quantum-classical correspondence. We start with the pure quantum eigenstates of the delta-kicked accelerator in the true semiclassical regime. The true semiclassical limit is achieved by letting the scaled Planck constant $\overline {k}=2 \pi\frac{T}{T_{1/2}}$ (see \cite{burnett}) approach zero, i.e. $T=\frac{1}{N} \; T_{1/2}$, for large $N$. Our aim is of course to compare the phase space of the classical states and  the quantum states. FIG.~\ref{q-cla}$(a)$ and $(b)$ show stroboscopic representations (termed Poincar\'e sections) of the relevant classical states. This representation is obtained by integrating Hamilton's equations over one period of the kicking potential. This gives rise, in terms of rescaled dimensionless variables, the following mapping: 
\begin{eqnarray}
\theta_{n+1}&=&\theta_n + \mathcal J_{n+1}\quad \text{mod}\left[ 2 \pi \right] \; ,\label{cmap1}\\ 
\mathcal J_{n+1} &=& \mathcal J_n - K \sin (\theta_n) - 2 \pi \Omega \quad \text{mod}\left[ 2 \pi \right] \; .  \label{cmap2}
\end{eqnarray} 
Here $K=2 \pi  k \frac{T}{T_{1/2}}$ is termed the stochasticity parameter, and $\Omega=GgT^2/2 \pi$. The parameters used in FIG.~\ref{q-cla} are $k=5$ and $T=\frac{1}{80} T_{1/2}$, i.e. $K \simeq 0.39$ and $\Omega = \frac{1}{2}$. In Fig.~\ref{q-cla}$(a)$, a classical accelerator mode of order $o=2$ and jumping index $j=1$ is observed: two KAM tori (i.e. regular trajectories) are centred around period-2 fixed points $A$ and $B$ of the mapping. These points define a period-2 periodic orbit $A \rightarrow B \rightarrow A' \rightarrow B'$ etc., which follows: $\mathcal J(A') = \mathcal J(A)-2 \pi$.
In FIG.~\ref{q-cla}$(c)$, we plot the Husimi function of a quasi-eigenstate (i.e. an eigenvector of the infinite matrix $\mathcal U$ which represents the evolution operator over one period of the kicking potential) for $\theta_0=0$. One should recall that $\theta_0$ is the analogue of a Bloch vector. We can, in this case, clearly identify the quantum state with its classical analogue as the KAM tori centred around the points $A$ and $B$ as associated with the accelerator mode. The quantum state is clearly localized on the invariant manifold constituted by the two KAM tori, and as these tori constitute a classical accelerator mode, we say that the quasi-eigenstates lie on this accelerator mode. FIG.~\ref{q-cla}$(b)$ show that 4 KAM tori are centred around a period-4 periodic orbit, and they belong to a classical accelerator mode of order $o=4$ and jumping index $j=2$. In FIG.~\ref{q-cla}$(d)$ we plot the Husimi function of a quasi-eigenstate which is clearly linked to the aforementioned period-4 periodic orbit. 
\begin{figure}[ht!] 
\begin{center}
\includegraphics[width=4cm,height=4cm,angle=0]{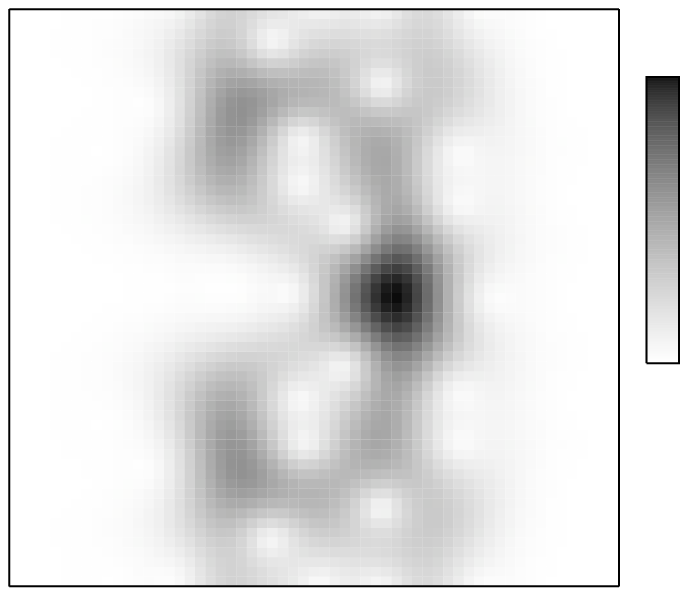}
\put(-105,55){$ \mathcal J$}
\put(-103,18){$0$}
\put(-107,92){$2 \pi$}
\put(-60,11){$\theta$}
\put(-97,11){$0$}
\put(-22,11){$2 \pi$}
\put(-60,102){$(a)$}
\put(-8,47){$0$}
\put(-8,83){$0.9$}
\put(-26,100){$|\langle \Phi \vert \theta , \mathcal J \rangle|^2 $}
\includegraphics[width=4cm,height=4cm,angle=0]{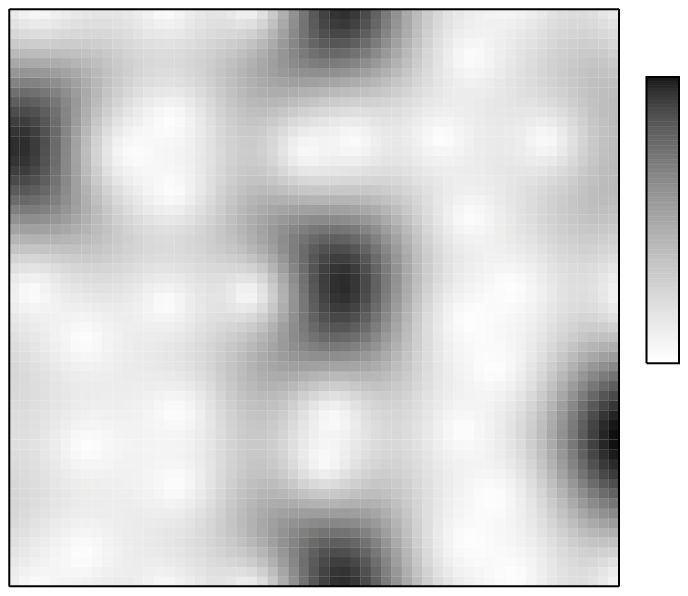}
\put(-60,102){$(b)$}
\put(-8,47){$0$}
\put(-8,83){$0.6$}
\caption{\small{The Husimi functions of two quasi-eigenstates for $T$ close to $T_{1/2}$. The parameters used are: $T=\frac{49}{50} \;T_{1/2}$, $\Omega=1$ in $(a)$ and $T=\frac{29}{30} \;T_{1/2}$, $\Omega=\frac{1}{2}$ in $(b)$, $\theta_0=0$, $k=1$. We do not represent the quantum phase-space cell in its entirety and limit ourselves to figures which can be compared with the $\epsilon$-classical phase-space. The quasi-eigenstate represented in $(a)$ might be linked to a period-$1$ fixed point of the $\epsilon$-classical map. $(b)$ could represent a quasi-eigenstate which lie on an $\epsilon$-classical accelerator mode.}}
\label{q-eps}
\end{center}
\end{figure} 
\\ \indent We will now study the quantum eigenstates of our system when the periodicity $T$ of the kicking potential is close to the resonant value $T_{1/2}$, $T=T_{1/2} (1 + \epsilon/2 \pi)$ with $\epsilon$ being small. In this regime, quantum accelerator modes are observed and a fraction of atoms are accelerated either faster or slower than the gravity alone would produce. This phenomenon is now purely quantum since there are no longer any classical accelerator modes in this regime. In fact the classical stochasticity parameter is so large that the whole phase-space cell is chaotic, and all the periodic orbits are unstable. However, quantum accelerator modes have close similarities with those of classical accelerator modes, i.e. they are like ``ghosts'' of the classical dynamic. In fact Fishman and al. have shown in \cite{FGR} that for $T$ close to $T_{1/2}$ the quantum dynamics follow the effective pseudo-classical map:
\begin{eqnarray}
{\theta}_{n+1} &=& {\theta}_{n} + \text{sgn}(\epsilon) {\mathcal J}_{n+1} \quad \text{mod}\left[ 2 \pi \right]\; , \quad  \\
{\mathcal J}_{n+1} &=& {\mathcal J}_{n} - K_{\epsilon} \sin\left( {\theta}_n  \right) - \text{sgn}(\epsilon) 2 \pi \Omega \quad \text{mod} \left[ 2 \pi \right] \; , \qquad 
\end{eqnarray}
(similar to Eq. $(\ref{cmap1})$ and $(\ref{cmap2})$). The pseudo-classical stochasticity parameter $K_{\epsilon}$ is defined as $K_{\epsilon}= \epsilon k$. This means that the smaller $\epsilon$, the more regular the $\epsilon$-classical map. Together with this shows that the $\epsilon$-classical accelerator modes account for the quantum accelerator modes. We now come to a crucial question we want to answer: are the quantum eigenstates related to these $\epsilon$-classical accelerator modes? Our examination of the Husimi functions of certain quasi-eigenstates leads us to conclude that the link is by no means a simple one. 
\begin{figure}
\begin{center}
\includegraphics[width=7cm,height=5cm,angle=0]{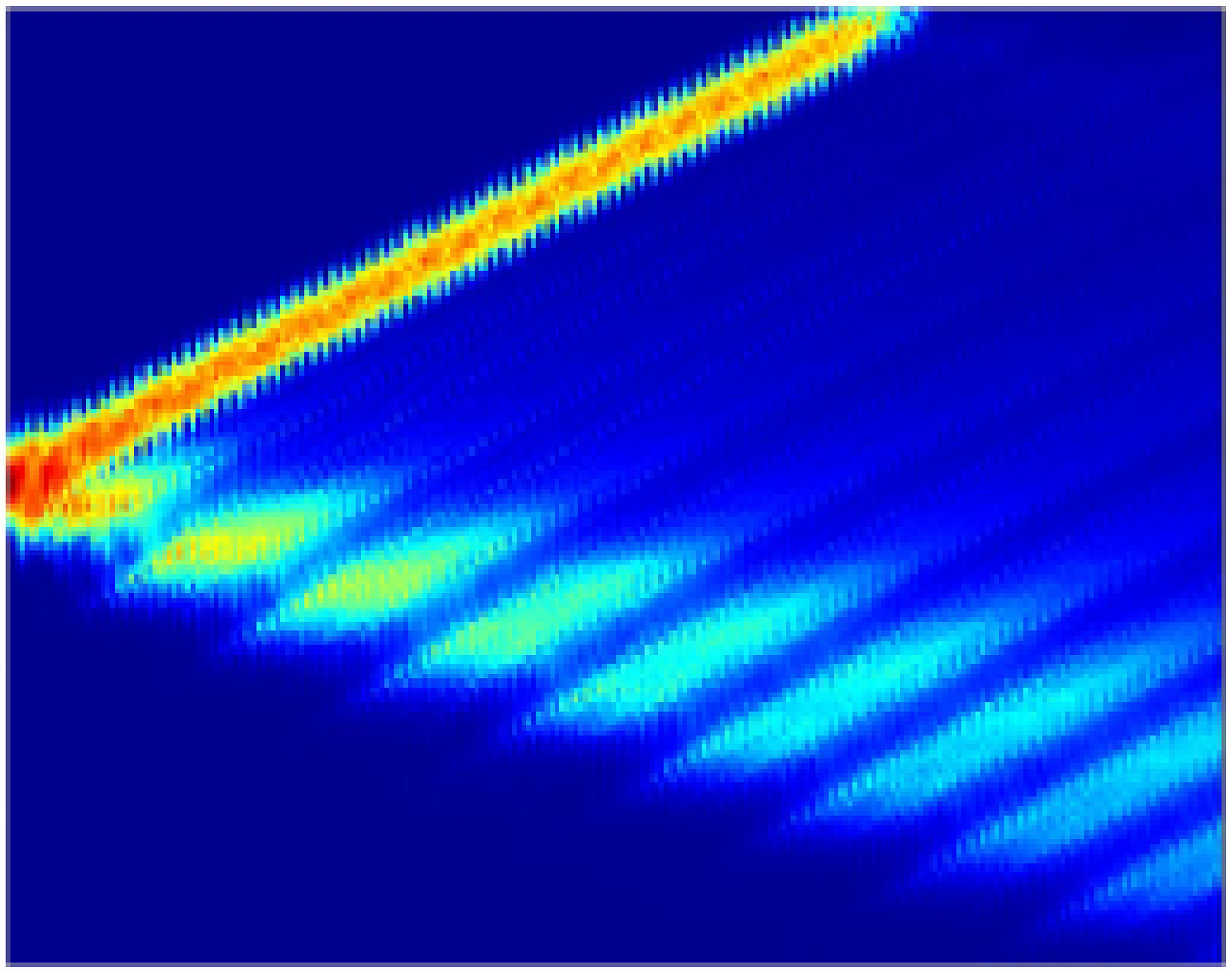}
\put(-100,138){$(a)$}
\put(-180,70){$0$}
\put(-187,125){$60$}
\put(-194,15){$-60$}
\put(-175,7){$0$}
\put(-25,7){$200$}
\put(-130,0){\texttt{Number of kicks}}
\put(-220,70){\texttt{$p$ [$ \hbar G]$}}
\end{center}
\begin{center}
\includegraphics[width=7cm,height=5cm,angle=0]{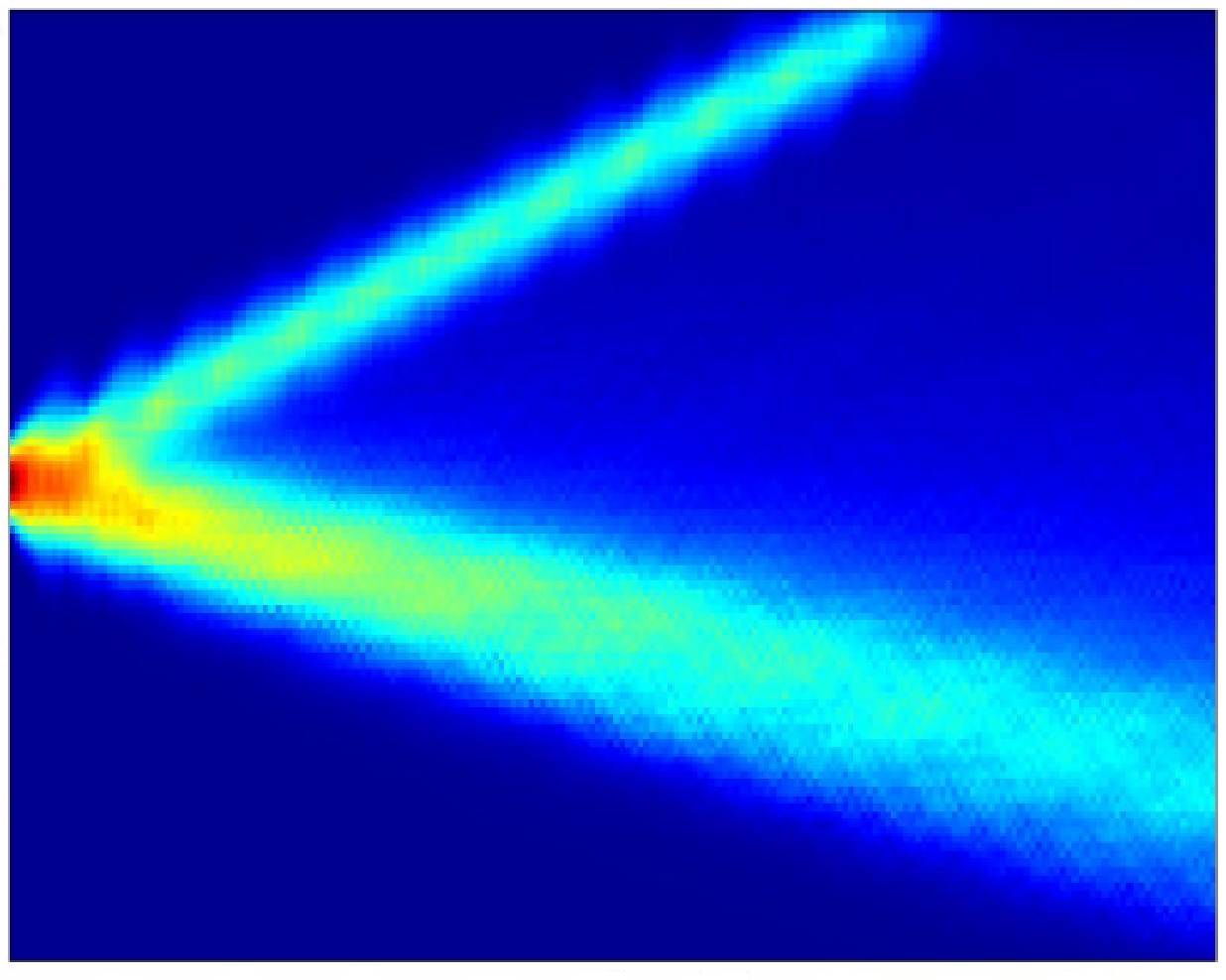}
\put(-100,138){$(b)$}
\put(-180,70){$0$}
\put(-187,125){$60$}
\put(-194,15){$-60$}
\put(-175,7){$0$}
\put(-25,7){$200$}
\put(-130,0){\texttt{Number of kicks}}
\put(-220,70){\texttt{$p$ [$\hbar G]$}}
\caption{\small{(color online) Numerical simulations for the quantum accelerator mode with $\Omega\simeq1/2$, $k=1$, in the semiclassical and $\epsilon$-semiclassical regimes. It shows the momentum variation with the kicking numbers of an initial state taken as an incoherent Gaussian mixture of plane waves with quasi-momentum $\beta$ centred around $\beta=0$. $(a)$ $T$ is small, $T=\frac{1}{20} T_{1/2}$, so that the scaled Planck constant is equal to $\overline {k}\simeq0.31$: we are in the semiclassical regime. The accelerator mode does not decay. $(b)$ $T$ is close to $T_{1/2}$, $T=\frac{19}{20} T_{1/2}$. The effective Planck constant $\epsilon$ is equal to $\epsilon=0.05$: we are in the $\epsilon$-semiclassical regime. The population accelerated decay: this is a quasi-accelerator mode.}}
\label{accelerator}
\end{center}
\end{figure} 
\\ \indent In FIG. \ref{q-eps}$(a)$ we show the Husimi function of a quasi-eigenstate when $T= T_{1/2} (1-1/50)$, $k=1$, $\Omega=1$, and $\theta_0=0$. We do not represent the quantum phase-space cell in its entirety and limit ourselves to a comprehensible figure, which can be compared with the $\epsilon$-classical phase-space. One cannot tell for sure whether this quasi-eigenstate is related to the $\epsilon$-classical period-$1$ fixed point, i.e. to the $\epsilon$-classical accelerator mode of order $o=1$ and jumping index $j=1$. The various peaks surrounding the main central Gaussian packet might be described by the cumulant theory introduced by Bach and al. \cite{burnett}. FIG. \ref{q-eps}$(b)$ represents the Husimi function of a quasi-eigenstate for $T= T_{1/2} (1-1/30)$, $k=1$, $\Omega=\frac{1}{2}$, and $\theta_0=0$. On the other hand, the function may be located on a period-$4$ periodic orbit. 
\\ \indent Our survey of the phase space leads us to conclude that the quantum phase-space cell do not contain structures related to accelerator modes. We can tentatively say that the quasi-eigenstate ``has escaped'' the accelerator mode in the relevant regions of the quantum phase-space cell.
\\ \indent It is of course the case that the $\epsilon$-classical accelerator modes~\footnote{We consider accelerator modes as invariant manifold of the evolution operator conjugated $o$-times with itself. Of course, the quasi-eigenstates are not associated with trajectories whose momentum increases.} are not necessarily the $\epsilon$-classical limit of Gaussian-smoothed Wigner functions representing pure quantum eigenstates. Indeed, they may represent mixtures of eigenstates associated with different quasi-energies, or ``quasi-accelerator modes'' that are reasonably long lived (of order $\epsilon^{-1}$ \cite{FGR}, see~\cite{Berry2} for a related work from M. V. Berry and al.) . In contrast to this, the accelerator modes which appear in the semiclassical regime do not decay (as long as the Hamiltonian (\ref{Hamiltonian}) accurately describes the dynamics of our system). Furthermore, the eigenstates which lie on classical accelerator modes remain constant in time and do not diffuse in either the position or the momentum space.\\  
\indent This is confirmed by the numerical simulation (see FIG.~\ref{accelerator}) for the quantum accelerator mode observed with the parameters taken as $\Omega=1/2$, $k=1$ in the semiclassical and $\epsilon$-semiclassical regimes. These simulations clearly show the difference between the true semi-classical regime and the pseudo-classical regimes.\\
\indent We wish to thank Zhao-Yuan Ma, D. Delande, B. Gr\'emaud, A. Voros, S. Nonnenmacher and J. Zyss for stimulating discussions and for bringing usefull references to our attention. G. Lemari\'e should like to thank A. Faurant for providing him computing resources. K. Burnett thanks the Royal Society and Wolfson foundation for support.

\end{document}